# X-ray detected ferromagnetic resonance spectrometer with an out-of-vacuum photodetector

T. Ueno[1,2,3], Y. Takeichi[4,3], M. Mizuguchi[5,2], H. Iwasawa[1,6], Y. Ohtsubo[6], K. Ono[3,4], H. Okazaki[7,1], S. Li[2,1], S. Sakai[2], T. Yamaki[7], T. Watanuki[1], Y. Katayama[1], and C. Mitsumata[8]

[1]Synchrotron Radiation Research Center, Kansai Institute for Photon Science, National Institutes for Quantum Science and Technology, Sayo, Hyogo 679-5148, Japan
[2]Quantum Materials and Applications Research Center, Takasaki Institute for Advanced Quantum Science, National Institutes for Quantum Science and Technology, Takasaki, Gunma 370-1292, Japan
[3]Department of Applied Physics, Graduate School of Engineering, The University of Osaka, Suita, Osaka 565-0871, Japan
[4]Institute of Materials Structure Science, High Energy Accelerator Research Organization, Tsukuba, Ibaraki 305-0801, Japan
[5]Institute of Materials and Systems for Sustainability, Nagoya University, Nagoya, Aichi 464-8603, Japan
[6]NanoTerasu Center, National Institutes for Quantum Science and Technology, Sendai, Miyagi 980-8579, Japan
[7]Department of Advanced Functional Materials Research, Takasaki Institute for Advanced Quantum Science, National Institutes for Quantum Science and Technology, Takasaki, Gunma 370-1292, Japan
[8]Graduate School of Pure and Applied Sciences, University of Tsukuba, Tsukuba, Ibaraki 305-8577, Japan

X-ray detected ferromagnetic resonance (XFMR) spectroscopy is an experimental technique for element-specific spin dynamics in the GHz regime and has been utilized to study spintronic materials. The XFMR signal is usually obtained by detecting X-ray excited optical luminescence (XEOL) emitted from a sample substrate. Here, we report the development of an XFMR spectrometer that is designed to place a photodetector for XEOL detection outside an ultra-high-vacuum chamber. This configuration allows for the easy replacement of detectors, such as photodiodes, CCD cameras, and spectrometers, depending on the experimental requirements. We demonstrated the measurement of XEOL spectra from MgO using a visible light spectrometer as well as the detection of XFMR signals originating from the spin precession of a permalloy ($Ni_{80}Fe_{20}$) thin film using a photodiode detector. The XFMR spectrometer with an out-of-vacuum photodetector expands possibilities for advanced measurements such as XFMR microscopy.

## 1. Introduction

In this era of information, the amount of electronic information is increasing rapidly. To process and store tremendous amounts of information in a sustainable manner, it is essential to improve the performance and energy efficiency of computing and storage devices. Spintronics is a promising technology that enables such advancements[1,2]. In addition to the electron charge degrees of freedom used in conventional electronics, spintronics also utilizes the electron spin degrees of freedom. Therefore, it is important to detect and control the electron spin. A wide range of materials are used in spintronic devices, including ferromagnetic[3], antiferromagnetic[4], and ferrimagnetic[5], as well as non-magnetic materials ranging from metals to insulators.

There are various experimental techniques for analyzing magnetic materials, but ferromagnetic resonance (FMR) is one of the most used. FMR measures the microwave absorption of a ferromagnetic sample placed in a static magnetic field. Recently, vector network analyzers (VNAs) have been widely utilized for FMR measurements[6]. By analyzing the FMR spectrum, magnetic properties such as effective magnetization, anisotropy, gyromagnetic ratio, damping, and exchange stiffness can be investigated. On the other hand, X-ray magnetic circular dichroism (XMCD) spectroscopy is a powerful experimental technique using synchrotron-radiation X-rays[7]. By tuning the energy of X-rays to the absorption edge of a specific element, element-selective magnetic measurements can be made. X-ray detected ferromagnetic resonance (XFMR) is a combined technique of FMR and XMCD spectroscopy that enables element-specific analysis of spin precession dynamics[8].

XFMR spectroscopy was demonstrated early this century and has been implemented in several synchrotron-radiation facilities[9–18]. This experimental technique has been applied to various kinds of magnetic systems such as single layer of films[9,11,19–21], exchange-coupled bilayers[13,14,22–28], and spin valves[29,30]. Spin-related phenomena such as spin pumping[31–38] and spin-transfer torque[39,40] have been observed. Leveraging the element-specific nature of XFMR, AC spin currents in non-magnetic layers[41–43] and spin waves in antiferromagnetic layers[44] have been observed. XFMR spectroscopy is realized not only with soft X-rays but also with hard X-rays[12,45,46]. Moreover, XFMR is combined with various X-ray experimental techniques such as

Corresponding author: T. Ueno (e-mail: ueno.tetsuro@qst.go.jp).

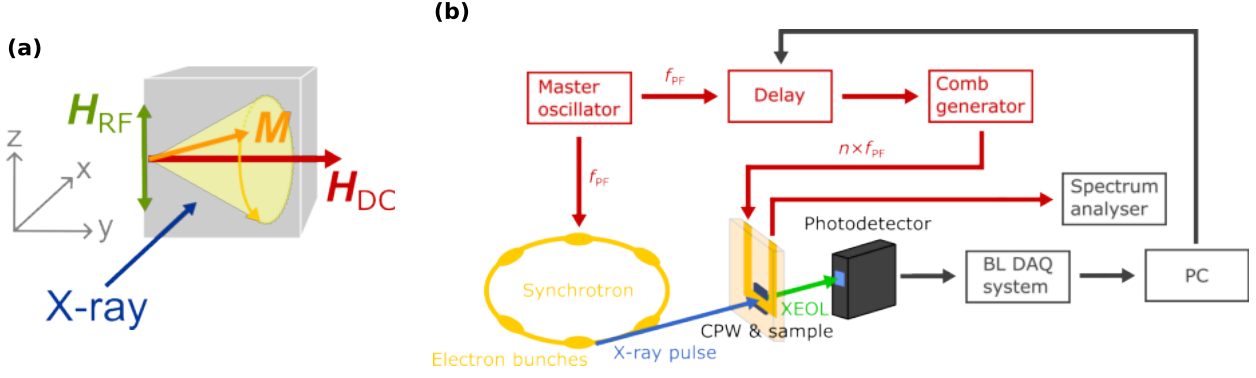

**Fig. 1** (a) Schematic of the X-ray detected ferromagnetic resonance (XFMR) spectroscopy experiment. Magnetization $M$ of a sample precesses along external static magnetic field $H_{DC}$ ($\parallel y$) under RF magnetic field $H_{RF}$ ($\parallel z$). Projection of $M$ along X-ray propagation direction ($\parallel x$) is detected via the XMCD effect. (b) Schematic of the present XFMR spectroscopy experimental system. $f_{PF}$ denotes frequency of the master oscillator of the Photon Factory ($f_{PF} = 500.1$ MHz). $n$ denotes an order of higher harmonics generated by a comb generator.

reflectometry[47], diffraction[48] and microscopy[49]. X-ray magnetic linear dichroism (XMLD) is also utilized as a probe for XFMR in ferrimagnetic materials[50]. Recent state-of-the-art XFMR measurements enable quantitative analysis of magnetic moment under ferromagnetic resonance[18,51].

Schematic of XFMR experiment is depicted in Fig. 1(a). Magnetization $M$ of a sample is aligned to the $y$ axis by external static magnetic field $H_{DC}$. RF magnetic field $H_{RF}$ with frequency $f_{RF}$ which is orthogonal to $H_{DC}$ is generated by CPW. Precession of $M$, termed as 'spin precession' hereafter, is induced when FMR conditions are met, i.e. appropriate combination of $H_{DC}$ and $H_{RF}$. XFMR signals are detected via the XMCD effect between circularly polarized X-ray and projection of $M$ along X-ray propagation direction ($x$ axis). Element-specific XFMR signal can be obtained by tuning X-ray energy at the specific absorption edges.

Spin precession is described with the Landau-Lifshitz-Gilbert (LLG) equation as follows,

$$\frac{dM_i}{dt} = -\gamma M_i \times H_{eff}^i + \alpha_i \left(\frac{M_i}{M_i} \times \frac{dM_i}{dt}\right),$$

where $M_i$, $H_{eff}^i$, and $\alpha_i$ are the magnetization, the effective magnetic field, and the Gilbert damping parameter of the specific element $i$, respectively. $M_i = |M_i|$ is the magnitude of the magnetization. $\gamma = g\mu_B/\hbar$ is the gyromagnetic ratio, where $g$ is the Landé g-factor and $\mu_B$ is the Bohr magneton. LLG equations of constituent elements are interpretable as simultaneous differential equations. The relation between the RF frequency $f_{RF}$ and the magnitude of the effective field $H_{eff}^i = |H_{eff}^i|$ is given by the Kittel equation as follows[52],

$$2\pi f_{RF} = \gamma \sqrt{H_{eff}^i(M_i + H_{eff}^i)}.$$

In this paper, we report design and performance of an XFMR spectrometer developed at the Photon Factory (PF), High Energy Accelerator Research Organization (KEK), Japan. In XFMR experiments, X-ray excited optical luminescence (XEOL) is usually detected to obtain an X-ray absorption signal. In most XFMR instruments, a photodiode at the sample manipulator under ultra-high vacuum (UHV) conditions is used to detect XEOL. This setup allows for a large numerical aperture (NA) to efficiently collect the XEOL signal emitted over a $4\pi$ solid angle; however, it restricts the quick replacement of detectors for specific experimental purposes. In the present XFMR spectrometer, a pair of plano-convex lenses collimates and focuses XEOL to the outside of the UHV chamber where photodetectors are placed. This design enables easy replacement of detectors, such as photodiodes with various wavelength sensitivities, CCD cameras, and spectrometers, depending on the experimental requirements.

## 2. Experimental

Synchrotron-radiation experiments were performed at BL-19B of the PF, Institute of Materials Structure Science, KEK, Japan. BL-19B is a branch beamline of BL-19A which is dedicated for scanning transmission X-ray microscopy (STXM)[53]. BL-19B is equipped with a bulk-XAS chamber that enables X-ray absorption spectroscopy (XAS) measurements of various types of samples using total electron yield (TEY), fluorescence yield, and transmission methods in the soft X-ray regime. A user can install a UHV chamber downstream of the bulk-XAS chamber, where we placed the XFMR spectrometer. BL-19A/B is equipped with an APPLE-II type undulator[54] and a Monk-Gillieson type monochromator[55]. Thus, soft X-ray with linear vertical and horizontal or left- and right-handed circular polarization is available. During the XFMR experiment, the PF was operated in multi-bunch mode. The electron beam bunch length in the storage ring and the corresponding X-ray pulse duration are approximately 35

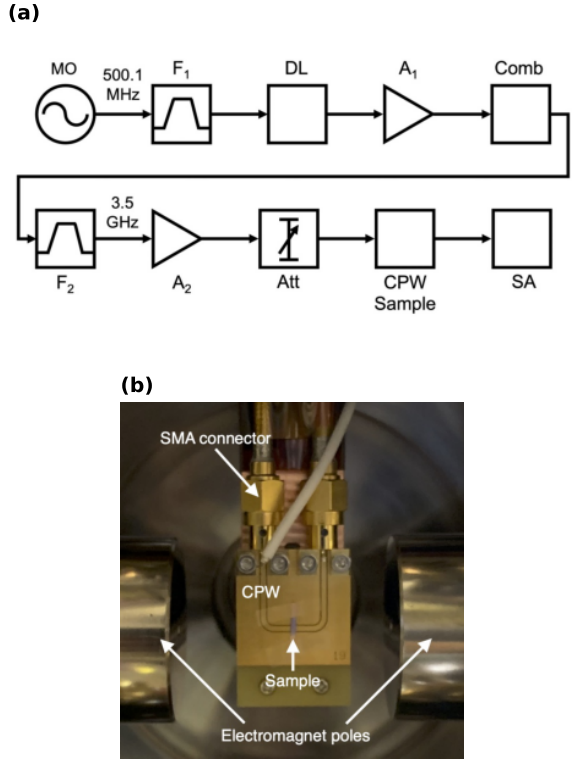

**Fig. 2** (a) Block diagram of the microwave circuit. Symbols indicate the master oscillator (MO), filters ($F_1$ and $F_2$), delay line (DL), amplifiers ($A_1$ and $A_2$), comb generator (Comb), attenuator (Att), coplanar waveguide (CPW), sample, and spectrum analyzer (SA). (b) Photograph of the CPW with a sample in the XFMR measurement configuration.

ps (rms)[56], which defines the temporal resolution of the XFMR experiment.

We used permalloy (Py), i.e., $Ni_{80}Fe_{20}$, thin films as samples. Py was deposited onto single-side-polished MgO(001) substrates at 300°C using magnetron sputtering. Two types of samples were prepared: (i) MgO(001)/Py(100 nm) and (ii) MgO(001)/Py(30 nm)/Cu(2 nm). A Cu capping layer was applied only to sample (ii) to prevent oxidation of the Py layer, whereas sample (i) was left uncapped to intentionally oxidize the Py film surface.

## 3. Details of the XFMR spectrometer

Schematic of the XFMR spectrometer is shown in Fig. 1(b). The experimental system mainly consists of two parts. One is a microwave circuit that generates higher harmonics of a frequency of the master oscillator to apply RF magnetic field to a sample. It also controls time delay between RF signals and X-ray pulses. The other is a XEOL detection optics and detector system, which is connected to the data acquisition (DAQ) system of the beamline.

### 3.1 Microwave circuit

Figure 2(a) shows a block diagram of the microwave circuit used in the XFMR spectrometer. The master oscillator (MO) signal of the PF ($f_{PF}$ = 500.1 MHz) was purified using a band-pass filter ($F_1$: 500 MHz band-pass filter, Mini-Circuits) and introduced to a delay line (DL: CDX-ATI011, Candox Systems Inc.) to control time delay between X-ray pulse and microwave signal into a coplanar waveguide (CPW). The delay line was driven by a motor driver (H718/GD-5410, Melec Inc.). Then, microwave signal was amplified by amplifier $A_1$ to drive a comb generator (Comb: Model 7102, Picosecond Pulse Labs Inc.) which generates higher harmonics of the input signal. A suitable combination of band-pass, low-pass, and high-pass filters ($F_2$: Mini-Circuits) allowed only the 3.5 GHz signal, which was used in this study, to pass through. Additional amplifiers ($A_2$: Mini-Circuits) and a programmable attenuator (Att: MN72A, Anritsu Co.) adjusted the power of the microwave signal. 3.5 GHz signal was introduced to CPW to generate RF magnetic field to drive spin precession in a sample. The transmitted microwave power was tuned to approximately 20 dBm, as measured by a spectrum analyzer (SA: FSV SIGNAL ANALYZER, Rohde & Schwarz GmbH & Co. KG).

To generate the RF magnetic field at the sample position, we used a two-port CPW with 50-ohm characteristic impedance. The CPW was designed and fabricated by HAYASHI-REPIC Co., Ltd. (Tokyo, Japan), with a signal line width of 1 mm. A 0.5-mm-diameter pinhole was fabricated at the center of the signal line to allow X-ray transmission. Figure 2(b) shows a photograph of the CPW with a sample in the XFMR measurement configuration. One port of signal line was connected to the RF signal generation system, which is described above, and another port was connected to the spectrum analyzer via SMA connectors. The sample, with its film side facing downward, was placed on the CPW so that it contacts the signal line, ensuring the effective application of the RF magnetic field.

### 3.2 XEOL detection

Figure 3(a) shows a schematic of the XEOL detection system from a side view. X-ray impinges on the film side of the sample through a φ5-mm pinhole on the signal line of the CPW. X-rays that have transmitted through the film excite XEOL in the substrate near the film-substrate interface. XEOL is emitted isotropically into a 4π steradian solid angle and is collected by a plano-convex lens (uncoated N-BK7 plano-convex lens LA1951, Thorlabs Inc.) mounted on lens tubes (vacuum-compatible SM1 lens tubes, Thorlabs Inc.) for collimation. The collimated XEOL then passes through another plano-convex lens (uncoated N-BK7 plano-convex lens LA1131, Thorlabs Inc.) and is focused onto the light-receiving element of a photodetector through an ICF34 viewport with borosilicate glass window. The detectable

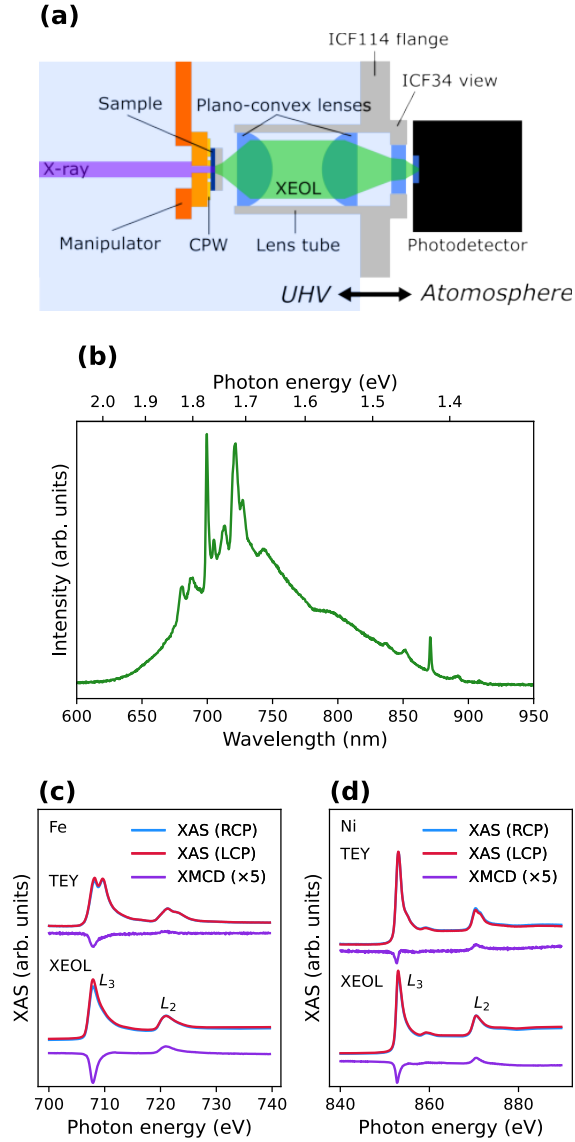

**Fig. 3** (a) Schematic of XEOL detection from the side view. (b) XEOL spectrum of MgO in visible light range. (c) Fe $L_{2,3}$ and (d) Ni $L_{2,3}$ XAS and XMCD spectra (magnified by a factor of 5 for visibility) of MgO(001)/Py(100 nm) measured by TEY and XEOL methods. RCP and LCP denote right- and left-handed circular polarization, respectively.

wavelength range of the optical system is determined by the material of these lenses and viewport window, and is 500–2000 nm, which includes wavelength of XEOL of substrates of interest.

In the present setup, an object lens with diameter φ = 25.4 mm and back focal length of 17.6 mm was adopted to achieve largest possible NA (NA ≈ 0.58). Note that the doublet lens system is in an UHV chamber and a photodetector is in atmosphere. Thus, in contrast to the previously reported XFMR experimental systems[8], a photodetector is installed outside the UHV chamber. Therefore, easy replacement of the photodetector is possible according to the purpose of the experiment, e.g. highly sensitive measurement with a photomultiplier tube detector and even microscopic imaging is possible with a CCD camera[57]. In the present setup, a photodiode module (C10439-03, Hamamatsu Photonics K.K.) was used. For light shielding, all viewports around the XFMR chamber except for XEOL optical path were closed, and the photodetector was enclosed with a blackout curtain.

Figure 3(b) shows XEOL spectrum of MgO measured with this system and a visible light spectrometer (FLAME-S, Ocean Insight, Inc.). Similar spectral shapes to those reported previously were observed, indicating $Cr^{3+}$ impurity plays a role as a luminous site in MgO[58].

To assess performance of the XEOL measurement system, we simultaneously measured Fe and Ni $L_{2,3}$ XAS of a 100-nm-thick Py film on MgO by TEY and XEOL as shown in Fig. 3(c)-(d). To measure XMCD, remanent magnetization was induced to the sample by applying in-plane magnetic field in advance. The sample film plane was set 9°-off-normal from the incident X-ray direction to enable observation of XMCD signal of projection of the remanent magnetization lies in the film plane. XMCD spectrum was obtained as a subtraction of XAS for right-handed circular polarization from that for left-handed circular polarization.

In TEY-detected Fe $L_{2,3}$ XAS spectra, two distinguishable peaks are observed at the Fe $L_3$ edge (upper side in Fig. 3(c)). The higher-energy side peak is originated from iron oxide at the film surface which is detected by surface-sensitive TEY method with a probing depth of about 2–3 nm. On the other hand, XEOL detection is bulk sensitive because film-transmitted X-ray excites XEOL in the substrate. Therefore, XEOL-XAS exhibits a single peak of metallic iron which comes from 100-nm-thick averaged signal (lower side in Fig. 3(c)). XMCD peaks appear only at the absorption edges of metallic iron both in TEY and XEOL detection. In the case of Ni $L_{2,3}$ XAS and XMCD spectra shown in Fig. 3(d), shoulder structures slightly appear at the higher energy side of the $L_{2,3}$ edges, however it is not obvious as in the case of the Fe $L_{2,3}$ edges. Overall spectral shapes of XEOL-XAS and XMCD are like those of bulk Ni including so-called 6-eV satellite[59]. As demonstrated above, bulk sensitivity of XEOL detection suppress relative contribution of oxidized surface layer and is well-suited for XFMR measurements. Moreover, XEOL detection enables one to probe buried magnetic layers in the depth which is hardly reached by TEY detection.

### 3.3 Performance of the XFMR spectrometer

To evaluate the performance of the XFMR spectrometer, we applied the system to MgO(001)/Py(30 nm)/Cu(2 nm) sample. Figures 4(a) and 4(b) show Fe and Ni $L_{2,3}$ XAS for circularly polarized X-ray. The $L_3$-edge energy values that exhibit the maximum XMCD amplitude were precisely determined in advance for the XFMR experiment. To establish the FMR condition, the static magnetic field $H_{DC}$ was adjusted to $H_{DC}$ = 15 mT

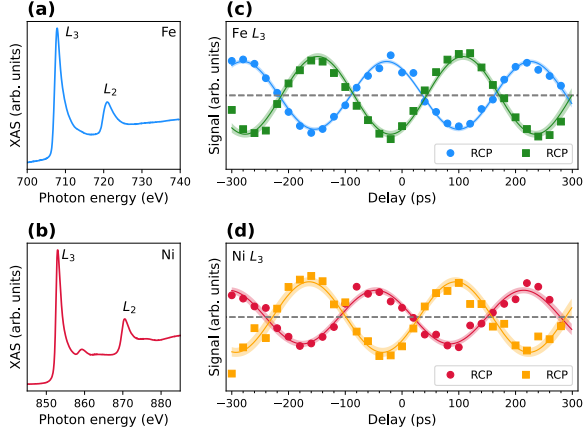

**Fig. 4** (a) Fe $L_{2,3}$ and (b) Ni $L_{2,3}$ XAS spectra of MgO(001)/Py(30 nm)/Cu(2 nm) measured by XEOL detection. Delay scan of MgO(001)/Py(30 nm)/Cu(2 nm) at (c) the Fe $L_3$ and (d) the Ni $L_3$ edges at $f_{RF}$ = 3.5 GHz. RCP and LCP denote right- and left-handed circular polarization, respectively. Solid curves represent the fitted curves, and shaded areas indicate the 95% confidence intervals.

while maintaining a fixed RF frequency of $f_{RF}$ = 3.5 GHz. The RF power absorption was monitored using a spectrum analyzer to optimize the magnetic field.

Figures 4(c) and 4(d) show delay scans in a 600 ps range at the Fe and Ni $L_3$ edges at the ferromagnetic resonance of 3.5 GHz, respectively. For both the Fe $L_3$ and Ni $L_3$ edges, the sinusoidal signals exhibit opposite signs for opposite circular polarizations of X-rays, confirming the successful observation of spin precession dynamics. Solid curves in Figs. 4(c) and 4(d) represent the sinusoidal fit of the experimental data. The fitted periods of the sinusoidal curves range from 253 to 260 ps, which are shorter than the expected period of $f_{RF}$ = 3.5 GHz (286 ps). We conclude that the period and phase obtained from the sinusoidal fits are within the expected temporal resolution of the experimental system. The phase of the sinusoidal signals is matched between Fe and Ni for the same X-ray polarization, indicating that the spins of both elements precess in phase within the Py film.

## 4. Discussion

Here, we discuss applications of the present XFMR apparatus equipped with out-of-vacuum photodetectors. As the schematic illustrated in Fig. 3(a), the photodetector can be easily replaced depending on the experimental purpose. For example, a two-dimensional (2D) detector, such as a CCD camera, can be employed instead of a point detector like a photodiode. By combining appropriate optics with a 2D detector, it is possible to realize X-ray excited optical microscopy (XEOM)[57,60] or X-ray excited luminescence microscopy (XELM)[61,62], which has the potential for spatially resolved XFMR measurements. Combined XEOM-XFMR measurements of composition-spread films fabricated via combinatorial methods could facilitate high-throughput characterization of element-specific magnetization dynamics in the GHz regime.

In X-ray absorption measurements using XEOL detection, the XEOL efficiency of the substrate material is a critical factor. This is particularly important for detecting weak signals, such as XFMR. The XEOL efficiency of MgO is comparatively higher than that of other transparent substrate materials, such as $Al_2O_3$ or $SrTiO_3$[63,64]. To enhance the efficiency of XEOL, we enriched $Cr^{3+}$ color center in MgO substrates by ion implantation[65] (not used in the present XFMR experiment). We successfully observed an enhanced XEOL signal from $Cr^{3+}$-implanted MgO, but further studies are required to optimize this approach.

## 5. Summary

We have developed the XFMR spectrometer with out-of-vacuum photodetectors. Details of the apparatus, including the microwave circuit and the XEOL detection system, have been presented. We demonstrated typical XFMR measurements, delay scans of a Py film on an MgO(001) substrate. A unique feature of this system is its out-of-vacuum photodetectors, which enables the integration of various functional detectors, such as a visible-light spectrometer or a CCD camera. This flexibility significantly expands the capabilities of XFMR, enabling measurements beyond conventional setups. The deployment of this type of XFMR spectrometer at a dedicated beamline in a high-brilliance, fourth-generation synchrotron radiation facility, such as NanoTerasu in Sendai, Japan, will provide new opportunities to observe element-specific spin dynamics in novel magnetic and spintronic materials and devices[66,67].

**Acknowledgements** This work was supported by the QST President's Strategic Grant (QST Advanced Study Laboratory) and the QST-Tohoku University Matching Research Support Program. TU acknowledges the support of JSPS KAKENHI (Grant Nos. JP15K17458 and JP18K13984), the Shimadzu Science Foundation, the Hyogo Science and Technology Association, and Murata Science and Education Foundation. MM acknowledges the support of JSPS KAKENHI (Grant No. JP21H05016). XFMR experiments were conducted at PF with the approval of the PF Program Advisory Committee (Proposal Nos. 2018MP001, 2021PF-G020, 2022G072, and 2024G565). The authors thank Christoph Klewe, Yuta Ishii, and Junichi Adachi for valuable comments on the XFMR technique and Shohei Yamashita for assistance with synchrotron radiation experiments.